\documentclass[article]{emulateapj}
\slugcomment{Draft}

\shorttitle{Tycho Stripes}
\shortauthors{Laming}

\begin{document}

\title{Wave Propagation at Oblique Shocks: How Did Tycho Get Its Stripes?}


\author{J. Martin Laming\altaffilmark1}


\altaffiltext{1}{Space Science Division, Naval Research Laboratory, Code 7684, Washington DC 20375
\email{laming@nrl.navy.mil}}

\begin{abstract}
We describe a new model for the ``stripes'' of synchrotron radiation seen in the remnant of Tycho's supernova. In our picture, cosmic rays streaming ahead of the forward shock generate
parallel propagating (with respect to the local magnetic field direction) circularly polarized
Alfv\'en waves that are almost free of dissipation, and due to being circularly polarized exhibit no spatial variation of magnetic field strength. Following interaction with the SNR shock with nonzero obliquity, these parallel propagating waves become obliquely propagating, due the the wave refraction (different in principle for the different plane wave components), and dissipation sets in. The magnetosonic polarization decays faster, due to transit time damping, leaving only the Alfv\'en mode. This surviving mode now exhibits a spatial variation
of the magnetic field, leading to local maxima and minima in the synchrotron emission, i.e. the stripes. We attribute the initial wave generation to the Bell instability, which in contrast to the resonant generation of upstream Alfv\'en waves, gives rise to a preferred wavelength, and hence the single wave period at which the stripes are seen. Based on estimates for damping rates due to turbulent cascade and transit time damping, we estimate the dependence of the
visibility of the stripes on the shock obliquity, and determine a maximum cosmic ray energy
in Tycho's SNR in the range $6\times 10^{14} - 1\times 10^{15}$ eV.
\end{abstract}

\section{Introduction}
The advent of the Chandra and XMM-Newton X-ray astronomy missions has revitalized the field
of cosmic ray acceleration. Chandra's combination of arcsecond resolution imaging and
the moderate energy spectral resolution afforded by CCD X-ray detectors has proven well
suited to the study of supernova remnants, and specifically the X-ray synchrotron radiation
emitted by cosmic ray electrons. This often reveals itself as thin rims of continuum
emission in the 4 - 6 keV waveband (a region essentially free of quasi-thermal line emission from shocked plasma) spatially coinciding with the supernova remnant forward shock. This can be understood \citep[e.g.][]{vink03}
as cosmic ray electrons radiating in the strong (i.e. amplified) magnetic field at the shock.
The radial extent is limited by either the radiative loss time of the electrons, or the
decay of magnetic field postshock. The remnant of Tycho's supernova shows an even more
intriguing pattern of synchrotron radiation; a series of ``stripes'' or ripples, \citep{eriksen11}, observed most clearly in the 4 - 6 keV waveband where
line emission from the quasi-thermal plasma in the SNR is absent. The wavelength corresponds to the gyroradius of
cosmic ray protons with energies in the range $10^{14} - 10^{15}$ eV, but the precise mechanism by which these structures form is not known.

The Bell instability \citep{bell04,bell05}, which amplifies magnetic field in the
shock precursor region, generates linearly polarized structures
in a near perpendicular shock geometry.
Considering the effect of these short wavelength fluctuations on the cosmic ray current,
\citet{vladimirov09} and \citet{bykov11a} have shown that long wavelength upstream structures can result, with spatial variations of the magnetic field strength and hence synchrotron emissivity. \citet{bykov11b} argue that these long wavelength structures are responsible for the ``stripes''. A number of conditions must be met. Most importantly, the
shock region where the stripes appear must be ``{\em nearly perpendicular}'' \citep[][don't specify how close to 90$^{\circ}$ they require]{bykov11b}, and that in this nearly perpendicular
region, shock acceleration must still be efficient. But as discussed elsewhere
\citep{zank06,laming13}, the efficiency of shock
acceleration at quasi-perpendicular shocks is open to question.

\citet{malkov12} offer an alternative idea, that is appears most promising at
parallel shocks. They find fully nonlinear exact ideal MHD solutions supported by
the cosmic ray return current, which comprise pulses of Alfv\'en waves that can propagate
ahead of the main shock. \citet{malkov12} argue that these Alfv\'en pulses when visible in X-rays should appear as quasi-periodic stripes, with spacing similar
to that of the observed stripes. In their equations, \citet{malkov12} neglect thermal and
cosmic ray pressure gradients, assuming that the ponderomotive force of the turbulence is
much stronger. In the opposite limit high frequency sound waves are generated by the Drury
instability \citep{Drury86}. \citet{caprioli13} suggest that the forward shock may push forwards in upstream cavities
created at the saturation of the Bell instability, also at parallel shocks. \citet{rakowski11}
discuss a similar idea in connection with shock structures (although not stripes) seen in SN 1006.

In this paper we pursue a different model for these synchrotron stripes in Tycho's SNR.
Cosmic rays drifting ahead of a quasi-parallel shock generate upstream circularly polarized Alfv\'en waves. We assume these to be parallel propagating \citep[see e.g.][]{bamert04,gargate12}, and with no spatial variation in magnetic field pressure.
Upon passage through the shock, both the magnetic field and the wave propagation change
direction, and by different amounts, so that the formerly parallel propagating waves are
now obliquely propagating. The circular polarization decomposes into its linearly polarized constituents. The magnetosonic polarization becomes compressive, and quickly decays by transit time damping. The Alfv\'en polarization survives longer, and its spatially varying magnetic field gives rise the regions of enhanced synchrotron emission, observable as the stripes or ripples
seen by \citet{eriksen11}. The following
sections treat the wave refraction at the shock, the wave transmission and reflection
coefficients, and the postshock damping of the magnetosonic and Alfve\'n modes. Section 5 puts these topics together to explain the origin of the stripes, and to derive physical implications
from this identification.

\begin{figure}[t]
\centerline{\includegraphics[scale=0.55]{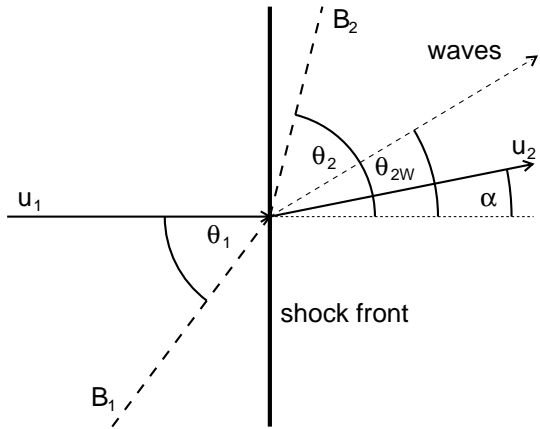}}
\caption{Schematic diagram of wave propagation at the oblique shock, in the shock rest frame. Upstream (left), plasma moves towards the shock front with velocity ${\bf u}_1$, carrying with it magnetic field ${\bf B}_1$ at angle $\theta _1$ to ${\bf u}_1$ (or the shock normal). Postshock, the magnetic field is ${\bf B}_2$ at angle $\theta _2$ to the shock normal and the flow velocity is ${\bf u}_2$ at angle $\alpha$ to the shock normal. In the preshock medium, waves are assumed to be parallel propagating along ${\bf B}_1$. Postshock,
the waves refract and travel at an angle $\theta _{2W}$ to the shock normal. They are no longer parallel propagating. In principle, Alfv\'en and magnetosonic polarizations will
refract at slightly different angles. The difference between these two angles of refraction tends to zero as the Alfv\'en Mach number $M_A\rightarrow\infty$, the approximation taken in this paper.\label{fig1}}
\end{figure}

\section{Wave Refraction at Shocks}
\citet{mckenzie70} and \citet{webb99} give formal accounts of wave properties at oblique shocks.
We follow and extend slightly for our particular application the more pedagogical treatment of \citet{achterberg86}.
A schematic diagram of the wave propagation at an oblique shock is shown in Fig \ref{fig1}.
In the preshock medium, we take both Alfv\'en and magnetosonic polarizations to be parallel
propagating (with respect to the upstream magnetic field) with phase velocity $V_{A1}$, the
upstream Alfv\'en speed. Downstream, the Alfv\'en polarization has phase speed $V_{A2}\cos\left(\theta _2-\theta _{2W}\right)$, where $\theta _2$ is the angle from the
shock normal to the postshock magnetic field, and $\theta _{2W}$ is the angle from the
shock normal to the wave propagation direction. The phase speed of the magnetosonic wave
depends on the plasma beta (the ratio of gas pressure to magnetic pressure),
and is given by \citep[e.g][]{melrose86}
\begin{eqnarray}
\nonumber v^2 &=&{1\over 2}\bigg(V_{A2}^2 + c_{s2}^2\pm\\
\nonumber & & \left[\left(V_{A2}^2+c_{s2}^2\right)^2-
4V_{A2}^2c_{s2}^2\cos ^2\left(\theta _2-\theta _{2W}\right)\right]^{1/2}\bigg)\\
\nonumber &\simeq &V_{A2}^2\cos ^2\left(\theta _2-\theta _{2W}\right)-{V_{A2}^4\over c_{s2}^2}\big[\cos ^2\left(\theta _2-\theta _{2W}\right)\\
& &
+\cos ^4\left(\theta _2-\theta _{2W}\right)\big] +\ldots\quad \beta >>1\\
& &  \simeq V_{A2}^2 + c_{s2}^2\sin^2\left(\theta _2-\theta _{2W}\right)\quad \beta <1.
\end{eqnarray}

In the following we will only consider the leading terms in equations 1 and 2, which will lead to the same angle of refraction $\theta _{2W}$ for both polarizations. In principle,
however, the different polarizations can refract at slightly different angles.
At the shock transition, we demand that the phases of upstream and downstream waves are equal, i.e.
$\omega _1t-{\bf k}_1\cdot {\bf r}=\omega _2t-{\bf k}_2\cdot {\bf r}$, and that the components of
wavevectors perpendicular to the shock velocity are continuous. For magnetosonic waves
at low $\beta$ these constraints give the equations
\begin{eqnarray}
& & k_1V_{A1}-k_1\cos\theta _1u_1 = k_2V_{A2}-k_2u_2\cos\left(\theta _{2W}-\alpha\right)\\
& & k_1\sin\theta _1 = k_2\sin\theta _{2W},
\end{eqnarray}
where $\tan\alpha = u_{2\perp}/u_{2\Vert} = ru_{2\perp}/u_{1\Vert}$, the tangent of the angle the postshock flow makes to the shock normal with $r$ being the shock compression ratio. Dividing (3) by (4) to eliminate $k_1$ and $k_2$ leads to
\begin{equation}
{V_{A1}\over \sin\theta _1} - u_1\cot\theta _1 = {V_{A2}\over\sin\theta _{2W}}-u_2\cot\theta _{2W}\cos\alpha
-u_2\sin\alpha ,
\end{equation}
which can be rearranged to give an equation for $\theta _{2W}$;
\begin{equation}
{M_{A2}\cos\theta _{2W} -1\over M_{A2}\sin\theta _{2W}} = r{M_{A1}\cos\theta _1-1\over M_{A1}\sin\theta _1}
-\tan\alpha .
\end{equation}
For $M_{A1}\rightarrow\infty$, $M_{A2}\rightarrow\infty$, (and hence $\tan\alpha\rightarrow 0$) and well away from the precisely perpendicular
shock where $\cos\theta _{2W}=\cos\theta _1=0$, this reduces to
\begin{equation}
\cot\theta _{2W}=r\cot\theta _1 = {r^2\over\tan\theta _2}
\end{equation}
where we have also used the result $\tan\theta _2=r\tan\theta _1$ from equation A4.
Writing $\tan\theta _{2W}=\tan\left(\theta _2-\Delta\right)$ where $\Delta$ is the angle between the magnetosonic wave propagation direction and the postshock magnetic field direction, we find
\begin{equation}
\tan\Delta = {\left(r^2-1\right)\over r}\cos\theta _1\sin\theta _1.
\end{equation}
The maximum deviation of the wave propagation from the magnetic field, and hence the maximum wave compression is achieved for $\theta _1 = 45^{\circ}$, where $\tan\Delta = -15/8$ (assuming $r=4$) and $\Delta = 62^{\circ}$. Solving
for $\theta _2$ we find $\theta _2=76^{\circ}$ and $\theta _{2W}=14^{\circ}$.

For magnetosonic waves at high $\beta$ and for Alfv\'en waves
equation 3 above becomes
\begin{eqnarray}
\nonumber & & k_1V_{A1}-k_1\cos\theta _1u_1 = k_2V_{A2}\cos\left(\theta _{2W}-\theta _2\right)\\
& & -k_2u_2\cos\left(\theta _{2W}-\alpha\right)
\end{eqnarray}
accounting for the different dispersion properties of the Alfv\'en wave. Following the same approach as above, we find
\begin{eqnarray}
\nonumber & & {M_{A2}-\cos\theta _2\over M_{A2}}\cot\theta _{2W}=r\cot\theta _1-{r\over M_{A1}\sin\theta _1}\\
& & -\tan\alpha
+{\sin\theta 2\over M_{A2}},
\end{eqnarray}
which can be recast as
\begin{eqnarray}
\nonumber & & \left(1-{r^{1/2}\cos\theta _1\over M_{A1}}\right)\cot\theta _{2W}=r\cot\theta _1-{r\over M_{A1}\sin\theta _1}\\
& & -\tan\alpha +{r^{3/2}\sin\theta _1\over M_{A1}}.
\end{eqnarray}
In the limit $M_{A1}\rightarrow \infty$, ensuring $\beta >>1$ in the postshock medium, this reduces to $\cot\theta _{2W}=r\cot\theta _1$ as above. Away from this limit, $\cot\theta _{2W}$ will be different for magnetosonic and Alfv\'en modes. In supernova remnants,
this is unlikely to be a significant effect, but in the lower $M_{A1}$ shocks driven by
solar coronal mass ejections, this could be an important consideration. The limit
$M_{A1}\rightarrow\infty$ also captures the case when $\omega _1\rightarrow 0$, appropriate
for the Bell nonresonant instability where modes grow at zero frequency.

\section{Wave Transmission and Reflection at Shocks}
\citet{vainio99} give transmission and reflection coefficients for forward and backward propagating Alfv\'en
waves at a parallel shock. Using the jump conditions for the tangential electric field,
\begin{equation}
\left[u_nB_t - B_nu_t\right]=0,
\end{equation}
the continuity of the transverse momentum,
\begin{equation}
\left[\rho u_nu_t - B_nB_t/4\pi\right]=0,
\end{equation}
and the continuity of the mass flux
\begin{equation}
\left[\rho u_n\right]=0,
\end{equation}
their result is
\begin{equation}
{T\atop R}= {\left(M_{A1}+H\right)\left(\sqrt{r}\pm 1\right)\sqrt{r}\over 2\left(M_{A1}\pm H\sqrt{r}\right)}.
\end{equation}
Here $M_A$ is the Alfv\'en Mach number of the shock, and $H$ is the cross helicity of the waves. $H=1$ for
forward and $H=-1$ for backward propagating waves. At a parallel shock, the two polarizations, Alfv\'en and magnetosonic, behave the same. At an oblique shock, differences emerge. For the Alfv\'en mode, which perturbs magnetic
field and velocity in the direction perpendicular to the plane in which the wave refraction occurs, results
for $T$ and $R$ are obtained from equation 15 with the replacement $M_{A1} \rightarrow M_{A1}/\cos\theta _1$.

The magnetosonic mode is much more involved since perturbed magnetic field and velocity vectors lie in the
plane of refraction. It appears to be tractable only in the limit $M_{A1}\rightarrow \infty$, when the motion of the shock front itself in response to the passing wave can be neglected. We evaluate the jump conditions
for the tangential electric field and the continuity of transverse momentum using
\begin{eqnarray}
& & B_n= -B_2\cos\theta _2 +\delta B\sin\theta _{2W}\\
& & B_t= B_2\sin\theta _2 +\delta B\cos\theta _{2W}\\
& & u_n=-u_{2\Vert} +\delta u\sin\theta _{2W}\\
& & u_t=u_{2\perp} +\delta u\cos\theta _{2W}.
\end{eqnarray}

Substituting into the first jump condition gives
\begin{eqnarray}
\nonumber & & \left(-u_1+\delta u_1^f\sin\theta _1\right)\left(B_1\sin\theta _1+\delta B_1^f\cos\theta _1\right)\\
& & -\left(-B_1\cos\theta _1+\delta B_1^f\sin\theta _1\right)\delta u_1^f\cos\theta _1\\
\nonumber & & =\left(-u_{2\Vert}+\delta u_2\sin\theta _{2W}\right)\left(B_2\sin\theta _2+\delta B_s\cos\theta _{2W}\right) \\
\nonumber& & -\left(-B_2\cos\theta _2+
\delta B_s\sin\theta _{2W}\right)\left(u_{2\perp}+\delta u_s\cos\theta _{2W}\right).
\end{eqnarray}
Multiplying out and assuming $-u_1B_1\sin\theta _1 = -u_{2\Vert}B_2\sin\theta _2-u_{2\perp}B_2\cos\theta _2$ yields
\begin{eqnarray}
\nonumber & & \delta u_1B_1-u_1\delta B_1\cos\theta _1=\delta u_2B_2\cos\left(\theta _2-\theta _{2W}\right)\\
& & -u_{2\Vert}
\delta B_2\cos\theta _{2W}-u_{2\perp}\delta B_2\sin\theta _{2W}.
\end{eqnarray}
We put $\delta u_1 = \delta u_1^f$ or $\delta u_1^b$ to represent an initially forward or backward propagating wave, and then put $\delta u_2 = \delta u_2^f +\delta u_2^b=\left(-\delta B_2^f+\delta B_2^b\right)/\sqrt{4\pi \rho _2}$ and rearrange to find
\begin{eqnarray}
\nonumber M_{A1}\cos\theta _1\pm 1&=&{T^f\atop R^b}\bigg\{{B_2\over \sqrt{r}B_1}\cos\left(\theta _2-\theta _{2W}\right)\\
\nonumber &+&{M_{A1}\over r}\cos\theta _{2W}+M_{A1}{u_{2\perp}\over u_1}\sin\theta _{2W}\bigg\}\\
\nonumber &-&{R^f\atop T^b}\bigg\{{B_2\over \sqrt{r}B_1}\cos\left(\theta _2-\theta _{2W}\right)-{M_{A1}\over r}\cos\theta _{2W}\\
&-&M_{A1}{u_{2\perp}\over u_1}\sin\theta _{2W}\bigg\}.
\end{eqnarray}
The +ve sign in the left hand side corresponds to $T_f=\delta B_2^f/\delta B_1^f$ and $R^f=\delta B_2^b/
\delta B_1^f$, while the -ve sign corresponds to $R^b=\delta B_2^f/\delta B_1^b$ and $T^b=\delta B_2^b/\delta B_1^b$.
A similar procedure for the second jump condition gives
\begin{eqnarray}
\nonumber \pm M_{A1}\cos\theta _1&+&\cos 2\theta _1={T^f\atop R^b}\bigg\{{M_{A1}\over\sqrt{r}}\cos\theta _{2W}\\
\nonumber &-&M_{A1}\sqrt{r}{u_{2\perp}\over u_1}\sin\theta _{2W}+{B_2\over B_1}\cos\left(\theta _2+\theta _{2W}\right)\bigg\}\\
\nonumber &+&{R^f\atop T^b}\bigg\{-{M_{A1}\over\sqrt{r}}\cos\theta _{2W}+M_{A1}\sqrt{r}{u_{2\perp}\over u_1}\sin\theta _{2W}\\
&+&{B_2\over B_1}\cos\left(\theta _2+\theta _{2W}\right)\bigg\}.
\end{eqnarray}

Taking $M_{A1}\rightarrow\infty$, so that also $u_{2\perp}\rightarrow 0$ in
equations 22 and 23 we find
\begin{eqnarray}
\nonumber {T^{bf}\atop R^{bf}}& = &{\cos\theta _1\over 2\cos\theta _{2W}}\sqrt{r}\left(\sqrt{r}\pm 1\right)\\
&=& \left(\sqrt{r}\pm 1\over 2\sqrt{r}\right)\sqrt{\sin ^2\theta _1 +
r^2\cos ^2\theta _1},
\end{eqnarray}
where we have substituted from $\cot\theta _{2W}=r\cot\theta _1$ in the final step. This agrees with \citet{vainio99} and equation 15 in the appropriate limits ($\theta _1\rightarrow 0$ and $M_{A1}\rightarrow\infty$).

The forgoing has treated Alfv\'en and magnetosonic waves of nonzero frequency in the upstream and downstream shock regions. As mentioned above, when being driven by the cosmic ray current, these waves grow in the upstream region at zero frequency, and hence $\delta u_1=\rightarrow 0$. In equations 22 and 23 this leads to the terms $\pm 1$ and $\pm M_{A1}\cos\theta _1$ on the left
hand sides of equation 22 and 23 respectively being dropped. We give the corresponding
expression for the transmission and reflection coefficients for Alfv\'en (A) and magnetosonic
(M) modes
\begin{eqnarray}
{T\atop R} &=& {M_{A1}r/\cos\theta _1\pm H\sqrt{r}\over2\left(M_{A1}/\cos\theta _1\pm H\sqrt{r}\right)}\rightarrow {r\over 2}\quad {\rm (A)}\cr
&=& {1\over 2}\sqrt{\sin ^2\theta _1 +
r^2\cos ^2\theta _1}\quad {\rm (M)},
\end{eqnarray}
which are independent of the cross-helicity $H$ in the limit that $M_{A1}\rightarrow\infty$.

\section{Wave Damping}
Behind the shock, the newly obliquely propagating waves are subject to various damping mechanisms. Both Alfv\'en and magnetosonic polarizations decay by turbulent cascade, and the magnetosonic wave, now compressive due to its oblique propagation also decays by transit time
damping by the shocked quasi-thermal ions.

Many formulations exist for the damping by turbulent cascade. We follow \citet{boldyrev05} who gives an expression to cover the cases of both the weak turbulence \citep{goldreich95} and
strong turbulence \citep{iroshnikov63,kraichnan65}. In these limits the approximate damping
rates for the large scale turbulence are
\begin{eqnarray}
\gamma _{GS}&=&k_{2\perp}\delta u_2=\omega\tan\left(\theta _2-\theta _{2W}\right)\delta u_2/V_{A2}\\
\gamma _{IK}&=&k_{2\perp}\delta u_2^2/V_{A2}=\omega\tan\left(\theta _2-\theta _{2W}\right)\delta u_2^2/V_{A2}^2,
\end{eqnarray}
where $\omega =k_{2\Vert}V_{A2}$ from equation 1.
The strong turbulence expression for $\gamma _{IK}$ is probably the most applicable.
These simplest expressions
refer to balanced turbulence, with equal wave amplitudes propagating in each direction.
In our case, and decay rate of the transmitted waves will depend on the intensity of counter
propagating reflected waves, and vice versa. We will assume cosmic rays streaming ahead of the shock generate wave travelling in one direction only (away from the shock), and then we will take $\delta u_2=\delta u_1R$, where $R$ is the reflection coefficient calculated above in equation 24, in the damping rates.

\begin{figure}[t]
\centerline{\includegraphics[scale=0.55]{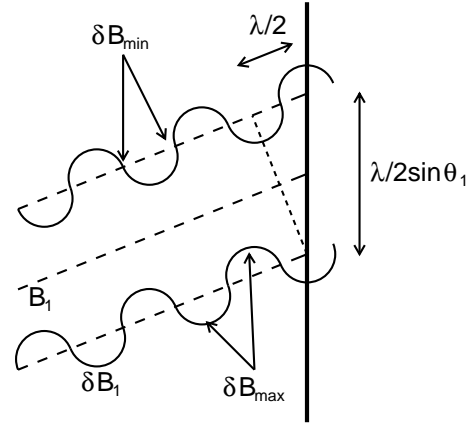}}
\caption{Schematic diagram of a plane parallel wave front encountering the shock. Upstream the wave is a parallel propagating circularly polarized wave, with no spatial variation in the
magnetic field strength. Behind the shock the magnetosonic polarization is damped quickly, leaving a spatial variation in magnetic field strength due to the surviving Alfv\'en polarization, with maxima every half wavelength along the wave propagation direction, $\lambda _1 /2$, as indicated.
When projected onto the shock front the distance between magnetic field maxima becomes
$\lambda _1/2\sin\theta _1$, where $\theta _1$ is the shock obliquity, the angle of the upstream unperturbed magnetic field to the shock normal.\label{fig2}}
\end{figure}

The magnetosonic polarization
is also subject to transit time damping by shocked but quasi-thermal ions. This damping rate in
the small gyroradius limit is given by \citep{melrose86}
\begin{eqnarray}
\nonumber & & \gamma _{TTD}=-\int\int {4\pi ^2q^2V_{A2}^2\over \hbar\omega c^2}\left(v_{\perp}^2k_{\perp}\over\Omega\right)^2\delta\left(\omega -k_{\Vert}v_{\Vert}
\right)\\
& & \times\hbar k_{\Vert}{\partial f\over\partial p_{\Vert}} 2\pi p_{\perp}dp_{\perp}dp_{\Vert},
\end{eqnarray}
where $\omega$ and $\Omega$ are the wave angular frequency and thermal ion gyrofrequency
respectively, $V_{A2}$ is the postshock Alfv\'en speed, and $v$ and $p$ are the ion velocity and momentum respectively, with subscripts $\perp$ and $\Vert$ indicating components
perpendicular or parallel to the ambient magnetic field. We represent the ion distribution function $f$ as a ``kappa'' distribution;
\begin{equation}
f={n\over \left(2\pi\right)^{3/2}m^3v_{Th2}^3}{\Gamma\left(\kappa\right)\over\Gamma\left(\kappa
-3/2\right)\kappa ^{3/2}}\left[1+{p^2\over 2\kappa ^2v_{Th2}^2}\right]^{-\kappa},
\end{equation}
where $m$ is the ion mass, $v_{Th2}$ is its thermal velocity, $\Gamma\left(\kappa\right)$ is
the Gamma function of with argument $\kappa$, and $\kappa$ is the index of the distribution.
For $\kappa\rightarrow\infty$, $f$ tends to a Maxwellian. For finite $\kappa$, the kappa
distribution has extended wings compared to a Maxwellian, and may be taken to represent
suprathermal ions in addition to the Maxwellian core. Normalizing the kappa distribution
over an infinite momentum range, we require $\kappa > 3/2$ to keep the number density $n$
finite, and $\kappa > 5/2$ to keep the energy finite. Smaller values of kappa require a
high-momentum cutoff to keep particle number and energy densities finite. Substituting
equation 29 into equation 28 we get
\begin{eqnarray}
\nonumber \gamma _{TTD}&=&{2\pi ^2q^2V_{A2}^2\over m^7v_{Th2}^5\Omega ^2c^2}{k_{\perp}^2\over k_{\Vert}}
\sqrt{2\over\pi}{n\Gamma\left(\kappa\right)\over\Gamma\left(\kappa -3/2\right)
\kappa ^{3/2}}\\
& & \times \left(2\kappa m^2v_{Th2}^2\right)^{\kappa +1}\\
\nonumber & & \times\int _0 ^{\infty}\left[2\kappa m^2v_{Th2}^2
+{m^2\omega ^2\over k_{\Vert}^2} +p_{\perp}^2\right]^{-\kappa -1}p_{\perp}^5dp_{\perp}.
\end{eqnarray}
The integral is performed by substituting $p_{\perp}=\sqrt{m^2\omega ^2/k_{\Vert}^2 +2\kappa m^2 v_{Th2}^2}\tan\psi$ and evaluating the resulting integral, $\int _0^{\pi /2}\sin ^5\psi
\cos ^{2\kappa -5}\psi d\psi=\int _0^1\left(1-\cos ^2\psi\right)^2\cos ^{2\kappa -5}\psi
d\left(\cos\psi\right)$ to find
\begin{eqnarray}
\nonumber & & \gamma _{TTD}={2\pi ^2q^2nV_{A2}^2\over m\Omega ^2c^2}{k_{\perp}^2v_{Th2}\over k_{\Vert}}
\sqrt{2\over\pi}\left[1+{V_{A2}^2\over 2\kappa v_{Th2}^2}\right]^{2-\kappa}\\
& & \times{\Gamma\left(\kappa\right)\over\Gamma\left(\kappa -3/2\right)
\kappa ^{3/2}}
{8\kappa ^2\over\left(\kappa -1\right)\left(\kappa -2\right)}.
\end{eqnarray}
As $\kappa \rightarrow\infty$, $\gamma _{TTD}\rightarrow 4\sqrt{2\pi }k_{\perp}^2v_{Th2}\exp\left({-1/\beta}\right)/k_{\Vert}$. For application below, where $\beta >>1$ and $\omega\simeq k_{\Vert}V_{A2}$, we rewrite as
\begin{equation}
\gamma _{TTD}\simeq 4\sqrt{2\pi }\omega\tan ^2\left(\theta _2-\theta _{2W}\right) v_{Th2}/V_{A2}.
\end{equation}
For a fast shock neglecting energy losses to cosmic rays, $v_{Th2}\simeq\sqrt{3}v_s/4\sim 2000$ km s$^{-1}$, where the shock velocity in Tycho $v_s\simeq 5000$ km s$^{-1}$.

\section{How Did Tycho Get Its Stripes?}
\citet{eriksen11} observe ``stripes'' in synchrotron emission from Tycho's SNR. The
stripes are separated by about 10'', which corresponds to $6\times 10^{17}$ cm at the
presumed 4.0 kpc distance \citep{hayato10}, or $5\times 10^{17}$ cm at 3.2 kpc \citep{slane14}.
We argue that these structures arise at oblique shocks, where upstream parallel circularly polarized waves which are undamped become
obliquely propagating damped waves postshock. The key point is that the magnetosonic polarization damps faster than the Alfv\'en polarization, and so a spatial variation in the magnetic field
strength will emerge associated with the surviving Alfv\'en mode. Along the direction of propagation the absolute magnitude of the magnetic field strength has a maximum every half
wavelength ($\lambda _1/2$). When projected onto the shock front, the maxima are separated by
a distance of $\lambda _1/2\sin\theta _1$, as shown schematically in Fig \ref{fig2}. Although less damped than the magnetosonic polarization, the Alfv\'en modes still damp within a wavelength or so postshock. These magnetic field variations are illuminated by synchrotron radiation from
cosmic ray electrons, the radiative cooling time for which typically lies between the magnetosonic and Alfv\'en mode damping times. This hierarchy ensures that the damped magnetosonic polarization does not contribute significantly to the emission, while the Alfv\'en mode only
contributes close to the shock, while the electrons are still radiating in X-rays.

\begin{figure}[t]
\centerline{\includegraphics[scale=0.45]{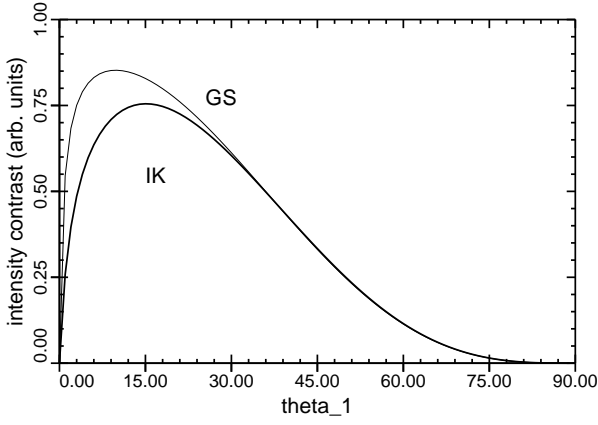}}
\caption{Variation with shock obliquity $\theta _{BN}$ of the intensity contrast $\Delta I$ between stripe maxima and minima, for the cases of Iroshnikov/Kraichnan (IK) or Goldreich-Sridhar (GS) turbulence.\label{fig3}}
\end{figure}

In the case of Iroshnikov-Kraichnan turbulence, the difference in synchrotron intensity between maxima and minima in the stripes may be written
\begin{eqnarray}
\nonumber \Delta I&\propto &\left[\exp\left(-\gamma _{IK}t\right)^{s+1} -
\exp\left(-\gamma _{IK}t-\gamma _{TTD}t\right)^{s+1}\right]\\
& & \times \cos ^{s+1}\theta _1
\end{eqnarray}
where $s$ is the index of the synchrotron photon spectrum and $s+1$ is the power law dependence
of the magnetic field strength on the synchrotron emission. The final factor of $\cos\theta _1$ gives the approximate shock obliquity dependence of the preshock magnetic field amplification.
Goldreich-Sridhar turbulence simply requires $\gamma _{GS}$ in place of $\gamma _{IK}$.
This intensity contrast is maximized at $t=\ln\left(1+\gamma _{TTD}/\gamma _{IK}\right)/\gamma _{TTD}/\left(s+1\right)$ with value
\begin{eqnarray}
\nonumber \Delta I&\propto &\bigg[\exp\left(-\gamma _{IK}/\gamma _{TTD}\right)\ln\left(1+\gamma _{TTD}/\gamma _{IK}\right)\\
\nonumber & & -\exp\left(-\gamma _{IK}/\gamma _{TTD}-1\right)\ln\left(1+\gamma _{TTD}/\gamma _{IK}\right)\bigg]\\
& & \times\cos ^{s+1}\theta _1.
\end{eqnarray}
Figure 3 shows the dependence of $\Delta I$ on the shock obliquity $\theta _1$, for both
Iroshnikov-Kraichnan (IK; strong) turbulence and Goldreich-Sridhar (GS; weak) turbulence,
calculated from equation 34, using equations 26, 27, and 32. We take $\delta u_2=\delta u_1R$,
and $\delta u_1\simeq 180$ km s$^{-1}$, corresponding to $R\simeq 2$ from equations 25, and postshock magnetic field and density of 180 $\mu$G and 1.2 amu cm$^{-3}$ respectively \citep{slane14}.
In both cases the emission is relatively strongly peaked over a restricted range of angles,
possibly suggesting a reason why the stripes are not ubiquitous over Tycho, but only seen in
certain ``special'' regions.

The separation at which the stripes preferably appear can be used to estimate the maximum cosmic
ray ion energy in Tycho. The Bell instability grows structures at a parallel wavevector in the ambient magnetic field given by \citep{bell04,bell05,laming14}
\begin{eqnarray}
\nonumber k_{\Vert}&=&1.5\times 10^{-8}\eta n_i\left(u_1\over 5000~{\rm km~s}^{-1}\right)^3
\left(3 \mu ~{\rm G}\over B\right)\\
& & \times {\gamma _{max}-3\gamma _1/4\over\gamma _{max}\gamma _1\left(\ln\gamma _{max}-1\right)}= {2\pi\over\lambda _1}.
\end{eqnarray}
Here $\eta$ is the fraction of shock energy going into cosmic rays, $n_i$ is the ion number density in the preshock medium. The highest energy cosmic rays have Lorentz factor $\gamma _{max}$, while the highest energy magnetized cosmic rays have Lorentz factor $\gamma _1$. The
cosmic ray current with particles with $\gamma _1 < \gamma <\gamma _{max}$ drives the Bell instability. We have also set $\delta B/B=1$ in equation 3 of \citet{laming14} for $k_{\Vert}r_g$, appropriate for the far upstream region of the cosmic ray precursor, and divided through by the cosmic ray gyroradius.

We identify $\lambda _1/2\sin\theta _1 = 5\times 10^{17}$ cm at a distance to Tycho of 3.2 kpc \citep[][model A]{slane14},
and taking $\theta _1$ from Fig 3, we estimate $\gamma _{max}$. Letting $\gamma _1\rightarrow\gamma _{max}$, and taking $\eta =0.26$, $n_i=0.3 $cm$^{-3}$, we find $\gamma _{max}= 6\times 10^5 - 1\times 10^6$ corresponding to $\theta _1=10^{\circ}-15^{\circ}$.  Although model dependent, this broadly agrees with the modeling of \citet{slane14}, who find the cosmic ray
energy spectrum breaking from something close to a $p^{-4}$ power law to a steeper fall-off
just above $\gamma\sim 10^5$ (their Fig. 4), and the original estimate of \citet{eriksen11},
who simply identified the spacing of the stripes with a cosmic ray gyroradius. Projected onto
the shock velocity vector, the wave phase varies with distance $\lambda _1/2 \times \cos\theta _1
\simeq 1\times 10^{17} - 1.5\times 10^{17}$ cm. Thus with a shock velocity of 5000 km s$^{-1}$,
changes in the projected position of the stripes should be visible in $2\times 10^8 - 3\times 10^8$ seconds, i.e. 6 - 10 years. \citet{eriksen11} report an intensity contrast of a factor of 25 between stripe maxima and minima. Relating this to $B^{s+1}$ with $s\sim 2.11$ we find a variation of $B$ between maxima and minima of about a factor 3. This is significantly lower than
the likely upstream magnetic field contrast of between a factor of 9 and 15, but probably reasonable when
we consider that both polarizations are amplified upstream by this amount, and then the
magnetosonic polarization is preferentially damped by transit time damping.
In the case that the electron cooling time becomes longer than the Alfv\'en mode
damping time, the variation of magnetic field along the shock front will be less clear, but
stripes separated by $\lambda _2/2$ oriented normal to ${\bf k}_2$ extending further inside the shock would result, analogous to ripples associated with sound waves.

\section{Conclusions}
We suggest a model for the synchrotron ``stripes'' or ripples observed in Tycho's SNR \citep{eriksen11} based only on simple ideas about Alfv\'en propagation and dissipation at
the forward shock, calculated as a function of its obliquity. We find that the contrast is highest, and therefore that the stripes would be most visible for a narrow range of shock
obliquities close to 15$^{\circ}$ in the case of strong Iroshnikov-Kraichnan turbulence. In our
model, the emission comes from behind the shock front, once the magnetosonic component of the
originally circularly polarized wave has been transit time damped. This is in contrast to previous ideas which have placed the structures giving rise to the synchrotron stripes in the
shock precursor. In the model advanced here, some motion of the stripes over a period of one to a few years should be visible, leading to possibilities for an observational test.

\acknowledgements This work was supported by Basic Research Funds of the Chief of Naval
Research. I acknowledge enlightening conversations with Chee Ng, and I am grateful to Una
Hwang and an anonymous referee for reading and commenting on the paper.

\appendix
\section{MHD Shock Relations}
Here we reproduce a few relationships relating to MHD shock, following mainly \citet{melrose86}.
The postshock perpendicular flow speed is
\begin{equation}
u_{2\perp}=u_1{\left(r-1\right)\sin\theta _1\cos\theta _1\over
M_{A1}^2-r\cos ^2\theta _1}\rightarrow 0\quad {\rm as}\quad M_{A1}\rightarrow\infty.
\end{equation}
The transverse magnetic field is
\begin{equation}
B_2\sin\theta _2={r\left(M_{A1}^2-\cos ^2\theta _1\right)\over M_{A1}^2-r\cos ^2\theta _1}B_1\sin\theta _1\rightarrow rB_1\sin\theta _1\quad {\rm as}\quad M_{A1}\rightarrow\infty.
\end{equation}
Writing
\begin{equation}
{B_2\sin\theta _2\over B_1\sin\theta _1}={r\left(M_{A1}^2-\cos ^2\theta _1\right)\over M_{A1}^2-r\cos ^2\theta _1}={\cos\theta _1\sin\theta_2\over\cos\theta _2\sin\theta _1}
\end{equation}
we find
\begin{equation}
\tan\theta _2=r\tan\theta _1+{\left(r-1\right)\sin\theta _1\cos\theta _1\over M_{A1}^2-r\cos ^2\theta _1} =r\tan\theta _1+{1\over r}\tan\alpha
\end{equation}
to be compared with
\begin{equation}
\cot\theta _{2W}=r\cot\theta _1 -\tan\alpha
\end{equation}
from equation 6, for the wave refraction.


\begin{thebibliography}{}
\bibitem[Achterberg \& Blandford(1986)]{achterberg86}Achterberg, A., \& Blandford, R. D. 1986, MNRAS, 218, 551
\bibitem[Bamert et al.(2004)]{bamert04}Bamert, K., Kallenbach, R., Ness, N. F., et al. 2004, \apj, 601, L99
\bibitem[Bell(2004)]{bell04}Bell, A. R. 2004, \mnras, 353, 550
\bibitem[Bell(2005)]{bell05}Bell, A. R. 2005, \mnras, 358, 181
\bibitem[Boldyrev(2005)]{boldyrev05}Boldyrev, S. 2005, \apj, 626, L37
\bibitem[Bykov et al.(2011a)]{bykov11a}Bykov, A. M., Osipov, S. M., \& Ellison, D. C. 2011a, \mnras, 410, 39
\bibitem[Bykov et al.(2011b)]{bykov11b}Bykov, A. M., Ellison, D. C., Osipov, S. M., Pavlov, G. G., \& Uvarov, Y. A. 2011b, \apj, 735, L40
\bibitem[Caprioli \& Spitkovsky(2013)]{caprioli13}Caprioli, D., \& Spitkovsky, A. 2013, \apj, 765, L20
\bibitem[Drury \& Falle(1986)]{Drury86} Drury, L.~O., \& Falle,
   S.~A.~E.~G.\ 1986, \mnras, 223, 353
\bibitem[Eriksen et al.(2011)]{eriksen11}Eriksen, K., Hughes, J. P., Badenes, C., et al. 2011, \apj, 728, L28
\bibitem[Gargat\'e \& Spitkovsky(2012)]{gargate12}Gargat\'e, L., \& Spitkovsky, A. 2012, ApJ, 744, 67
\bibitem[Goldreich \& Sridhar(1995)]{goldreich95}Goldreich, P., \& Sridhar, S. 1995, \apj, 438, 763
\bibitem[Hayato et al.(2010)]{hayato10}Hayato, A. et al. 2010, \apj, 725, 894
\bibitem[Iroshnikov(1963)]{iroshnikov63}Iroshnikov, P. 1963, Astron. Zh., 40, 742
\bibitem[Kraichnan(1965)]{kraichnan65}Kraichnan, R. H. 1965,Phys. Fluids, 8, 1385
\bibitem[Laming et al.(2014)]{laming14}Laming, J. M., Hwang, U., Ghavamian, P., \& Rakowski, C. E. 2014, \apj, 790, 11
\bibitem[Laming et al.(2013)]{laming13}Laming, J. M., Moses J. D., Ko, Y.-K., Murphy, R. J., Ng, C. K., Rakowski, C. E., \& Tylka, A. J. 2013, ApJ, 770, 73
\bibitem[Malkov et al.(2012)]{malkov12}Malkov, M. A., Sagdeev, R. Z., \& Diamond, P. H. 2012, \apj, 748, L32
\bibitem[McKenzie \& Westphal(1970)]{mckenzie70}McKenzie, J. F., \& Westphal, K. O. 1970, Phys. Fluids, 13, 630
\bibitem[Melrose(1986)]{melrose86}Melrose, D. B. 1986, Instabilities in Space and Laboratory Plasmas (Cambridge: Cambridge University Press)
\bibitem[Rakowski et al.(2011)]{rakowski11}Rakowski, C. E., Laming, J. M., Hwang, U., Eriksen, K. A., Ghavamian, P., \& Hughes, J. P. 2011, \apj, 735, L21
\bibitem[Slane et al.(2014)]{slane14}Slane, P. O., Lee, S.-H., Ellison, D. C., Patnaude, D. J., Hughes, J. P., Eriksen, K. A., Castro, D., \& Nagataki, S. 2014, \apj, 783, 33
\bibitem[Vainio \& Schlickeiser(1999)]{vainio99}Vainio, R. \& Schlickeiser, R. 1998, \aap, 331, 793
\bibitem[Vink \& Laming(2003)]{vink03}Vink, J., \& Laming, J. M. 2003, \apj, 584, 758
\bibitem[Vladimirov et al.(2009)]{vladimirov09}Vladimirov, A. E., Bykov, A. M., \& Ellison, D. C. 2009, \apj, 703, L29
\bibitem[Webb et al.(1999)]{webb99}Webb, G. M., Zakharian, A., Brio, M., \& Zank, G. P. 1999, J. Plasma Physics, 61, 295
\bibitem[Zank et al.(2006)]{zank06}Zank, G. P., Li, G., Florinski, V., Hu, Q., Lario, D., \& Smith, C. W. 2006, J. Geophys. Res. Space Phys., 111, 6108
\end{thebibliography}
\end{document}